\documentclass[prb,onecolumn,preprintnumbers,amsmath,amssymb,11pt]{revtex4-2}
\usepackage[utf8]{inputenc}
\usepackage{epsf}
\usepackage{graphicx}
\usepackage{latexsym}
\usepackage{amsmath}
\usepackage{amssymb}
\usepackage{amsfonts}
\usepackage{color}

\usepackage{enumitem}

\usepackage{xcolor}
\definecolor{darkblue}{rgb}{0.00, 0.00, 0.55}
\definecolor{darkmagenta}{rgb}{0.50, 0.00, 0.50}
\definecolor{darkcerulean}{rgb}{0.03, 0.27, 0.49}

\usepackage{amsmath}
\usepackage{amssymb}
\usepackage{graphicx,epsf}
\usepackage{hyperref}
\hypersetup{
    colorlinks         = true,
    citecolor          = darkblue,
    linkcolor          = darkblue,
    filecolor          = magenta,
    urlcolor           = darkmagenta,
    bookmarks             = false,
    unicode                  = true,
    bookmarksopen       = true,
    bookmarksopenlevel = 1
}

\begin{document}

\title{\Large{Effective removal of global tilt from topography images of vicinal surfaces with narrow terraces}}

\author{A. Yu. Aladyshkin$^{1-3, *}$, A. N. Chaika$^{4}$, V. N. Semenov$^{4}$, A. M. Ionov$^{4}$, S. I. Bozhko$^{4}$}

\medskip
\affiliation{$^1$Institute for Physics of Microstructures Russian Academy of Sciences, GSP-105, 603950 Nizhny Novgorod, Russia\\
$^2$Center for Advanced Mesoscience and Nanotechnology, Moscow Institute of Physics and Technology, Institutsky str. 9, 141700 Dolgoprudny, Russia\\
$^3$Lobachevsky State University of Nizhny Novgorod, \mbox{Gagarin Av. 23, 603022 Nizhny Novgorod, Russia} \\
$^4$Osipyan Institute for Solid State Physics Russian Academy of Sciences, \mbox{Acad. Osipyan str. 2, 142432 Chernogolovka, Russia} \\}


\maketitle



\section*{Abstract}

The main feature of vicinal surfaces of crystals characterized by the Miller indices $(h\,h\,m)$ is rather small width (less than 10 nm) and substantially large length (more than 200 nm) of atomically-flat terraces on sample surface. This makes difficult to apply standard methods of image processing and correct visualization of crystalline lattices at the terraces and multiatomic steps.  Here we consider two procedures allowing us to minimize effects of both small-scale noise and global tilt of sample: (i) analysis of the difference of two Gaussian blurred images, and (ii) subtraction of the plane, whose parameters are determined by optimization of the histogram of the visible heights, from raw topography image. It is shown that both methods provide nondistorted images demonstrating atomic structures on vicinal Si(5\,5\,6) and Si(5\,5\,7) surfaces.

$^*$ Corresponding author, e-mail address: aladyshkin@ipmras.ru

\section{Introduction}

Scanning tunneling microscopy (STM) is unique experimental technique providing detailed information about local structural and electronic properties of surfaces of solids \cite{Binnig-82,Chen-93,Voigtlaender-15}. However raw topography images are usually distorted due to global tilt of sample, nonorthogonality of fast- and slow-scanning directions of piezo-scanner, creep of piezo ceramics leading to noncontrolled drift of scanning area etc. As a consequence, raw topography images require justified corrections. Global tilt of the sample with respect to the reference coordinate system can be routinely removed by subtraction of an approximate plane, parameters of which can be determined by considering any of atomically-flat terraces of rather large size. Distortions in the lateral directions  caused by creep and thermal drift can be compensated afterwards provided that scanning area incorporated some fragments of crystalline lattice with well-known structure like Si(1\,1\,1)\,$1\times 1$, Si(1\,1\,1)\,$7\times 7$ or graphite honeycomb lattice. Such the correction procedures are incorporated in commercial and free-distributed applications like WsXM \cite{wsxm-ref}, Gwyddion \cite{gwyddion-ref} and others.

\begin{figure}[t!]
    \centering
    \includegraphics[width=85 mm]{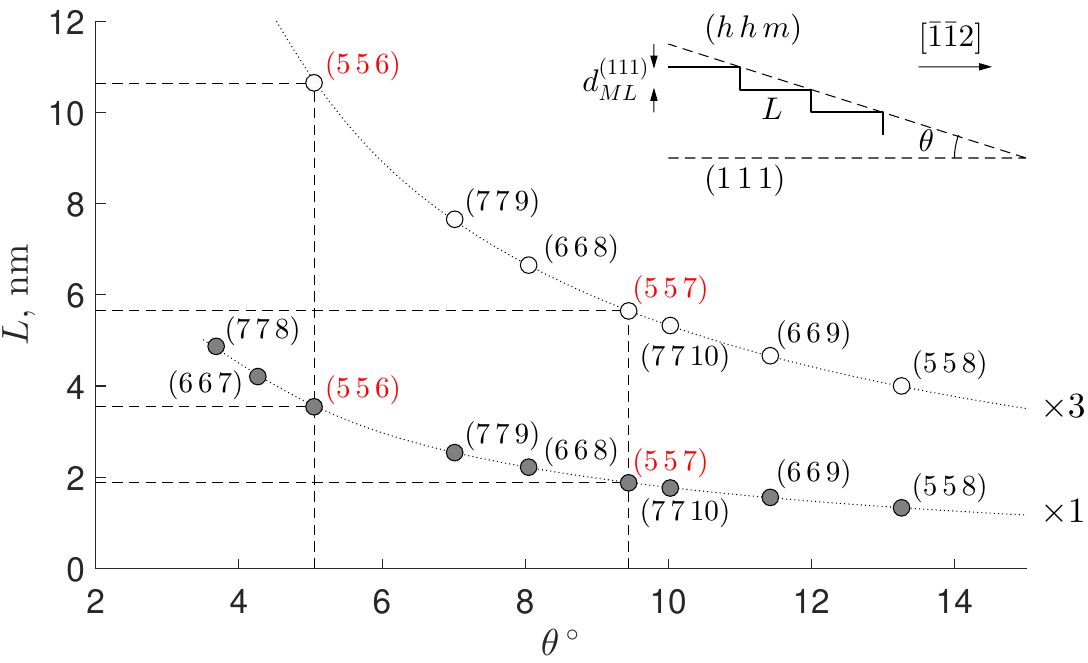}
    \caption{Dependence of the equilibrium width $L$ of atomically-flat Si(1\,1\,1) terraces between monatomic steps ($\bullet$) and triple steps ($\circ$) on the angle $\theta$ between planes $(h\,h\,m)$ and $(1\,1\,1)$ for several types of vicinal Si surfaces [see Eq.~(\ref{Eq-width-of-terrace})]. Inset illustrates the relationship between the terrace width $L$ and the interlayer distance $d^{(111)}_{ML}=a/\sqrt{3}= 0.314\,$nm, where $a=0.543$\,nm is the unit cell for bulk Si crystals.
    \label{Fig-01-Miller-planes}}
    \end{figure}

This paper is devoted to the problem of reliable elimination of the global tilt for samples with vicinal surfaces. Vicinal or high-Miller-index stepped surface \cite{Oura-03} is crystalline surface, which is misoriented from closely-packed low-Miller-index plane by a small angle $\theta$ (see Fig.~\ref{Fig-01-Miller-planes} and  comment \cite{Comment}). As a result, such surface can be viewed as a periodic combination of atomically-flat low-Miller-index terraces and mono- or multiatomic steps between them. Vicinal Si($h\,h\,m$) surfaces seem to be perspective semiconducting substrates for fabrication of ordered low-dimensional systems with unique electronic structure, transport and magnetic properties \cite{Erwin-10,Losio-01,Ahn-PRL-03,Lipton-Duffin-08, Tegenkamp-05,Morikawa-10,Brand-15,Quentin-20,Nita-22}. Such array of terraces and steps can be considered both as natural cheap and reproducible quantum standards of distances and heights and as perspective substrates for deposition of ultra-thin metallic films and design of quasi-one-dimensional metallic wires \cite{Losio-01,Ahn-PRL-03,Lipton-Duffin-08,Tegenkamp-05,Morikawa-10,Brand-15,Quentin-20}. It was demonstrated \cite{Kirakosian-APL-01} that the formation of atomically-precise ordered array of Si(1\,1\,1)\,$7\times 7$ terraces and quantized steps, whose heights are equal to three interplane distances (so called triple steps), is possible on Si(5\,5\,7) surfaces with typical lateral dimensions of the order of $100\,$nm.

The main problem in processing topography images of clean vicinal surfaces is related to the fact that atomically-flat terraces are rather narrow, what introduces substantial uncertainty in estimating the parameters of the approximate plane and thus complicates accurate elimination of the global tilt using standard procedures. Indeed, the period of a series of identical multiatomic steps can be calculated using the following relationship (see inset in Fig.~\ref{Fig-01-Miller-planes})
\begin{eqnarray}
\label{Eq-width-of-terrace}
L = \frac{n\,d^{\,}_{ML}}{{\rm tg}\,\theta},
\end{eqnarray}
where $d^{\,}_{ML}$ is the interplane distance (or thickness of a monolayer) for a family of nearest closely-packed terraces, $n$ is the multiplicity factor. The dependence $L$ on $\theta$ for several types of vicinal surfaces of Si single crystals is presented in Fig.~\ref{Fig-01-Miller-planes}. It is easy to see that the widths of the (1\,1\,1) terraces between monatomic steps on vicinal surfaces Si(5\,5\,6) and Si(5\,5\,7) are quite small (3.55\,nm and 1.88\,nm, respectively).

Despite the standard procedure of subtraction of the plane from raw topography image for Si single crystals with narrow terraces is badly conditioned, some authors have successfully applied this method for visualization of atomic structure of vicinal surfaces \cite{Chaika-APL-09}. Some authors seemingly considered misaligned topography images expecting that the presence of tilted terraces on topography images does not affect their scientific conclusions \cite{Teys-06,Zhachuk-09,Zhachuk-14,Chaika-SurfSci-09,Kim-10,PerezLeon-16,PerezLeon-17,Oh-08}.
Kirakosian \emph{et al.} \cite{Kirakosian-APL-01} have applied procedure of numerical differentiation of topography signal along the fast-scanning direction to visualize atomic structure on step-like Si(5\,5\,7) surface. However, this approach apparently converts raw topography data into a picture with distorted contrast, which cannot be referred to actual crystalline lattice. Numerical differentiation of the STM images was also used in other studies \cite{Chaika-JAP-09}. In order to balance contrast along the multiatomic steps, P\'{e}rez Le\'{o}n \emph{et al.} \cite{PerezLeon-17} have subtracted linear fit for each line in raw topography image. Apparently, this method works only if the step edges on the rotated topography image are parallel to horizontal direction.

\medskip
Some methods suitable for the considered problem were developed by scientists working in fields of image processing/recognition and computer vision (see, \emph{e.g.}, Ref.~\cite{Gonsales-04}). The approach known as the difference-of-Gaussians \cite{Gonsales-04,Kovasznayl-53,Marr-80,Lindeberg-15} is frequently used for automatic detection of defects, lines and edges on halftone and pseudocolor images. Recently this approach was used for electron microscopy \cite{Krivanek-15,Misra-20} and atomic-force microscopy \cite{Marsh-15} image processing. To the best of our knowledge, this simple and handy method was never applied for problems of scanning tunneling microscopy in general and for the processing of topography of vicinal surfaces in particular. One of our goals is to introduce this well-known method to the surface-science community dealing with complicated topography images of non-flat surfaces with steps, facets, bubbles etc (see also \cite{Aladyshkin-submitted}). To illustrate better capabilities of the difference-of-Gaussians approach we present the analysis of raw topography images of vicinal Si(5\,5\,6) and Si(5\,5\,7) surfaces with one-dimensional arrays of triple steps \cite{Kirakosian-APL-01,Chaika-APL-09,Teys-06,Zhachuk-09,Zhachuk-14,Chaika-SurfSci-09,Oh-08,Chaika-JAP-09,Henzler-03}.

The second part of the paper is devoted to the description of a histogram-based method of alignment of raw topography images. The idea is quite simple: one can determine the proper parameters of the plane, which should be subtracted from raw data, by minimizing the widths of the peaks for the probability density function $f(z)$, characterizing the statistical distribution of the visible heights for partly aligned images. We demonstrate that this algorithm can be easily realized in any programming language. This approach allows us to analyze both two-dimensional periodicity for atomically-flat terraces and precisely estimate the heights of multiatomic steps. We hope that this new method will be convenient for automatic analysis of large datasets of raw topography images.

\section{Experimental procedure}

Experimental investigations of crystalline structures of vicinal Si(5\,5\,6) and Si(5\,5\,7) surfaces were carried out in an ultrahigh vacuum (UHV) scanning tunneling microscope GPI-300 ($\Sigma-$scan) operating at a base vacuum pressure of \mbox{$1\cdot 10^{-10}\,$mbar}. All STM measurements were performed at room temperature in the regime of constant tunneling current $I$ and fixed electrical potential of the sample $U$ with respect to a grounded probe. Electrochemically etched poly- and single crystalline W tips were used as STM probes after \emph{in situ} electron bombardment and ion etching in UHV chamber \cite{Chaika-SciRep-14}. Freely-distributed applications WsXM \cite{wsxm-ref} and Gwyddion \cite{gwyddion-ref} as well as original programs written by authors in Matlab and Python were used for processing of topography images.

    \begin{figure}[h!]
    \centering
    \includegraphics[width=85 mm]{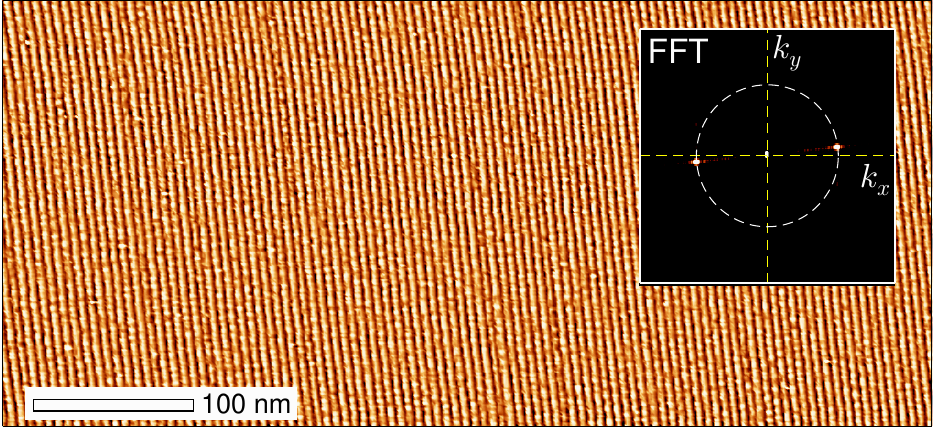}
    \caption{Large-scale topography image of Si(5\,5\,7) surface, confirming the formation of periodic one-dimensional (1D) array of triple steps (image size of \mbox{$583\times 267\,$nm$^2$}, tunneling voltage of \mbox{$U=-1.80\,$V}, mean tunneling current of \mbox{$I=50\,$pA}). Inset shows the structure of the $k$-space obtained by fast Fourier transform (image size \mbox{$4\times 4\,$nm$^{-2}$}). The radius of the circle is equal to $2\pi/L=1.11\,$nm$^{-1}$, where $L\simeq 5.65\,$nm is the mean width of the Si(111) terraces between ideal triple steps for the Si(5\,5\,7) surface  (see Fig.~\ref{Fig-01-Miller-planes}).
    \label{Fig-02}}
    \end{figure}

Si samples with typical dimensions of 0.5\,mm$\times 3\,$mm$\times 8$\,mm were made from polished $n$-type single-crystal Si(5\,5\,6) and Si(5\,5\,7) wafers (P-doped, resistivity of the order of $25\,\Omega\cdot$cm at 300\,K). These samples were first outgassed at moderate temperatures ($\sim 600-650^{\circ}$C) for 15-20 hours inside UHV chamber in order to remove contaminations from sample and contact plates without damaging natural oxide layer at the sample surface. For preparing clean surfaces with periodic step array we used the procedure of \emph{in situ} direct-current annealing (current is oriented perpendicular to steps and it flows in the step-up direction). This procedure allows us to reproducibly fabricate periodic array of triple steps at the sample surface of large (micron-size) areas (Fig.~\ref{Fig-02}). Note that prolonged annealing at higher temperatures at the final stage leads to substantial increase in widths of atomically-flat Si(1\,1\,1) terraces and formation of macrosteps in some surface areas similar to that described by Latyshev and Aseev \cite{Latyshev-98}.

\section{Results and discussion}

\subsection{Extraction of periodic signal associated with surface reconstructions}

We start with a trivial remark that the convolution of an arbitrary function $f(x)$ with a Gaussian function
\begin{eqnarray}
\label{Eq-def-gaussian}
G^{\,}_{\sigma}(x) = \frac{1}{\sqrt{2\pi}\sigma}\,\exp\left(-\frac{x^2}{2\sigma^2}\right)
\end{eqnarray}
plays pole of a low-pass filter, which effectively weakens Fourier components in the spectrum of $f(x)$ with large $k$ values exceeding certain threshold value of the order of $1/\sigma$, where $\sigma$ is the width of the Gaussian filter. In other words, the Gaussian blurring suppresses small-scale noisy components without affecting medium-scale and large-scale components.

Let us consider a function of two variables $z(x,y)$, which can be associated with raw topography image, acquired by scanning tunneling microscopy. Generally speaking, this function can have (i) small-scale noisy component, (ii) periodic two-dimensional (2D) component due to a presence of crystalline lattice, and (iii) large-scale inhomogeneities due to global tilt of the sample, multiatomic steps on surface, internal stress etc.

    \begin{figure}[t!]
    \centering
    \includegraphics[width=70 mm]{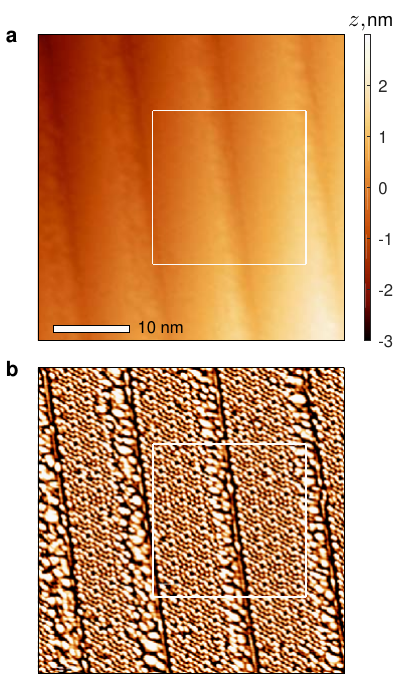}
    \caption{\textbf{a} -- Raw topography image of Si(5\,5\,6) surface (image size of \mbox{$40\times 40\,$nm$^2$}, $U=-0.60\,$V, $I=80\,$pA). \textbf{b} -- Map of the differential signal $D(x,y)$, visualizing the crystalline lattice at atomically-flat Si(1\,1\,1) terraces. This image is obtained from the raw image on the panel a by means of Eq.~(\ref{Eq-diff-gaussians}) without additional image processing; smoothing parameters are equal to $\sigma^{\,}_1=0.055$\,nm (0.5\,pxl) and $\sigma^{\,}_2=5\sigma^{\,}_1$. The area within the square depicted by solid white line is also shown in Fig.~\ref{Fig-04-Comparision-diff-width}.
    \label{Fig-03}}
    \end{figure}

We introduce the difference of two blurred images prepared by the convolutions of raw topographic image $z(x,y)$ and two Gaussian functions of various widths $\sigma^{\,}_1$ and $\sigma^{\,}_2>\sigma^{\,}_1$
\begin{multline}
\label{Eq-diff-gaussians}
D(x,y) =  \iint  \frac{z(x',y')}{2\pi \sigma^2_1}\,e^{-\left((x-x')^2+(y-y')^2\right)/2\sigma^2_1}\,dx'dy' - \\ \iint \frac{z(x',y')}{2\pi \sigma^2_2}\,e^{-\left((x-x')^2+(y-y')^2\right)/2\sigma^2_2}\,dx'dy'.
\end{multline}
It is easy to show that procedure (\ref{Eq-diff-gaussians}) corresponds to band-pass filtering in the $k$-space and it is similar to numerical calculation of the Laplacian of the blurred function. \cite{Aladyshkin-submitted} This difference-of-Gaussians method is frequently used in image processing for automatic detection of edges, defects etc.\cite{Gonsales-04,Kovasznayl-53,Marr-80,Lindeberg-15} We apply this approach to get rid of noisy component and global slope in topographical images and thus to highlight both periodic 2D corrugation associated with surface reconstructions on all terraces. In addition, this method makes possible to study irregular atomic structure at multiple steps which would be hidden in usual topography STM images because of large step height substantially exceeding typical atomic corrugations at steps and terraces. The strategy of choosing the parameters $\sigma^{\,}_1$ and $\sigma^{\,}_2$ is rather simple: $\sigma^{\,}_1$ should be of the order of pixel size, while $\sigma^{\,}_2$ should be larger than $\sigma^{\,}_1$ but smaller than typical length scales of structural inhomogeneities.

    \begin{figure}[h!]
    \centering
    \includegraphics[width=90 mm]{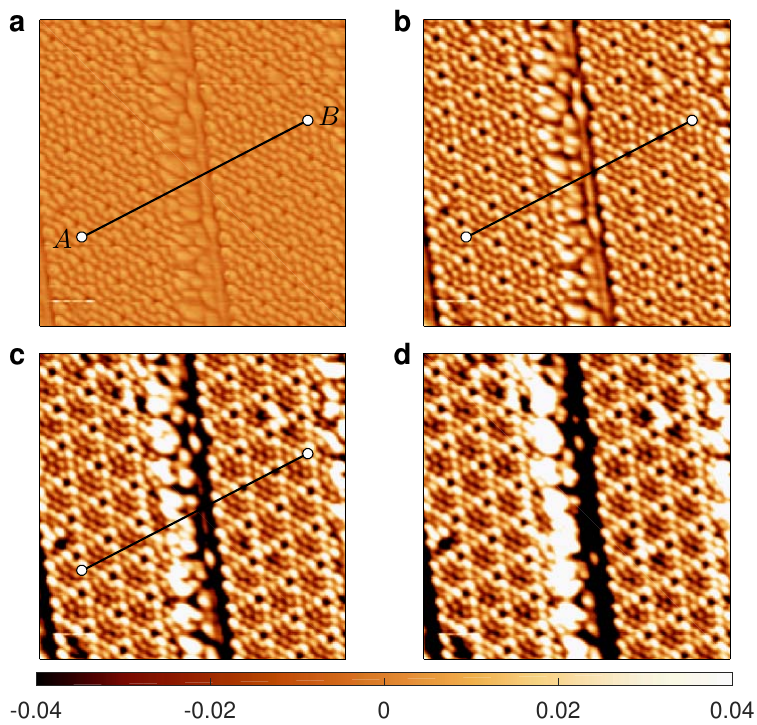}
    \includegraphics[width=90 mm]{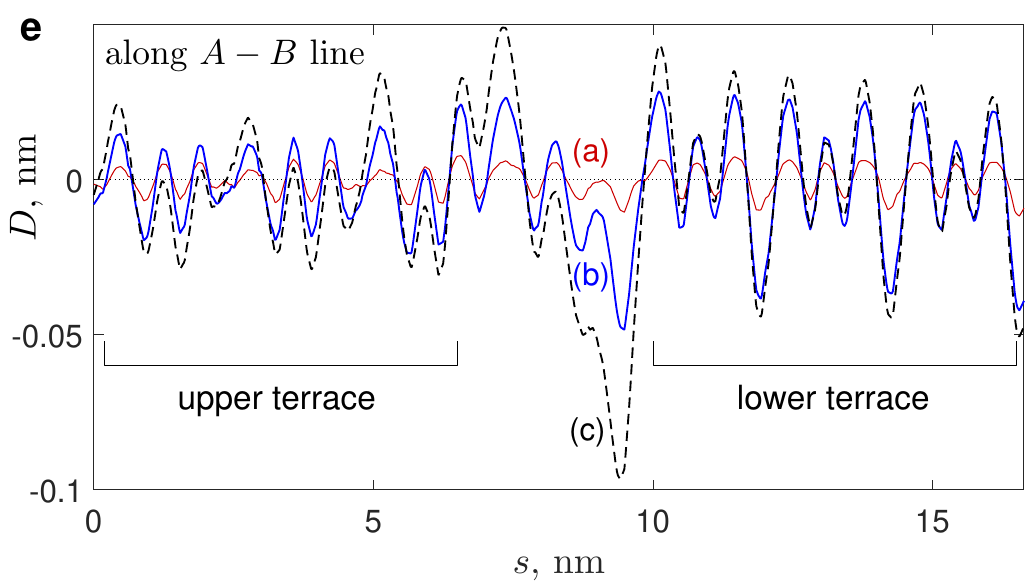}
    \caption{\textbf{a}--\textbf{d} -- Maps of the differential signal $D(x,y)$, composed for  the inner part of the topography image in Fig.~\ref{Fig-03}a (image size of \mbox{$20\times 20\,$nm$^2$}) for the different smoothing parameters: $\sigma^{\,}_1=0.055$\,nm (0.5 pixel) and $\sigma^{\,}_2=2\sigma^{\,}_1$ (panel a), $5\sigma^{\,}_1$ (panel b), $10\sigma^{\,}_1$ (panel c) and $15\sigma^{\,}_1$ (panel d). Note that all these images are composed using the same color scale (see horizontal color bar in the bottom). \textbf{e}~-- Cross-sectional views, composed for the maps in the panels a, b and c, along the $A-B$ line.
    \label{Fig-04-Comparision-diff-width}}
    \end{figure}

    \begin{figure}[t!]
    \centering
    \includegraphics[width=60 mm]{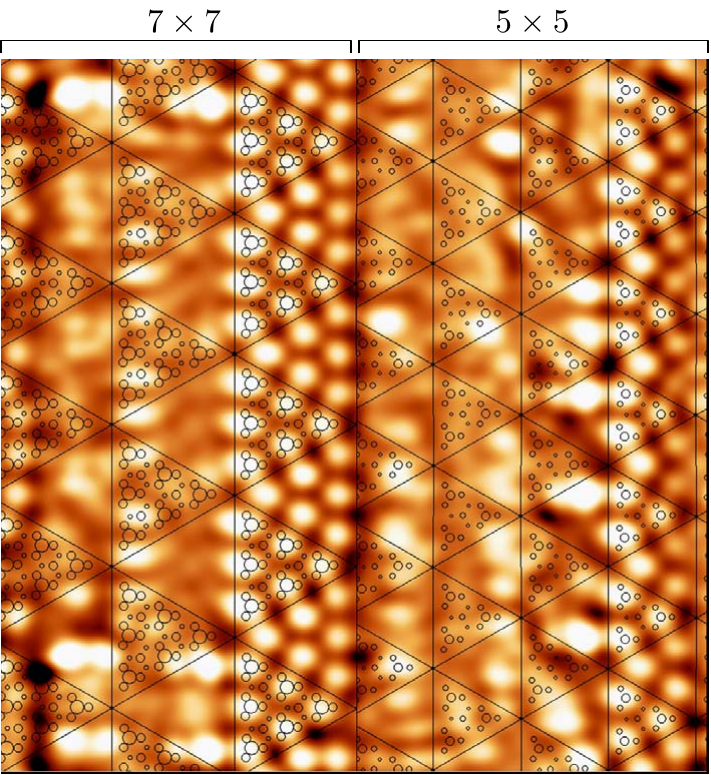}
    \caption{Map of the differential signal $D(x,y)$ for Si(5\,5\,7) surface (\mbox{image size of $13.4\times 13.4\,$nm$^2$}, $U=1.50\,$V, $I=70\,$pA) after affine transformation eliminating the drift of piezo-scanner. We plot crystalline structures \mbox{Si(111)$7\times 7$} (left part of the image) and \mbox{Si(111)$5\times 5$}  (right part of the image) on top of the $D(x,y)$ map.
    \label{Fig-05-Chaika}}
    \end{figure}

Examples of extraction of the periodic components, associated with 2D reconstructions on the Si(1\,1\,1) terraces, from raw topography images for the Si(5\,5\,6) surface by means of the difference-of-Gaussians procedure (\ref{Eq-diff-gaussians}) are presented in Figs.~\ref{Fig-03}b and \ref{Fig-04-Comparision-diff-width}. Since the observed period of the steps on the Si(5\,5\,6) surface is close to 10\,nm (Fig.~\ref{Fig-03}b), we conclude that these steps should be triple ones.

We would like to emphasize that the procedure (\ref{Eq-diff-gaussians}) is rather robust and it seems to be not very sensitive to particular values of the smoothing parameters $\sigma^{\,}_1$ and $\sigma^{\,}_2$. Indeed, increasing in the $\sigma^{\,}_2$ value (for the fixed $\sigma^{\,}_1$ parameter) leads to change in contrast for the map of the differential signal without modification of the periodic components related to the $7\times 7$ reconstruction on the terraces and the periodicity of the step array (Fig.~\ref{Fig-04-Comparision-diff-width}). We can only mention that changing the $\sigma^{\,}_2$ to $\sigma^{\,}_1$ ratio one can enhance/lessen various details in topography images that can be practically used for direct visualization and precise analysis of STM data obtained on vicinal surfaces. For example, the differential signal obtained at $\sigma^{\,}_2$ to $\sigma^{\,}_1$ ratios equal to 2:1 and 5:1 (Fig.~\ref{Fig-04-Comparision-diff-width}a,b) can be used for simultaneous visualization of the terrace and step atomic rows with comparable contrast despite the large height difference related to the presence of multiple steps on the surface. This cannot be achieved using plane or other standard correction procedures generally used for STM image processing. Contrary to typical differentiated STM images (\emph{e. g.}, $\partial z/\partial x$), maps of the differential signal $D(x,y)$ reproduce well all details of the atomic structure with exactly the same positions of maxima and minima as it could be observed on perfectly aligned topography image (Figs.~\ref{Fig-03}b and \ref{Fig-04-Comparision-diff-width}). Thus, the difference-of-Gaussians maps provide correct visualization of topography images projected onto scanning $x-y$ plane irrespective to particular large-scale tilt in various areas of considered surface.

For Si($h\,h\,m$) surfaces with $7\times 7$ reconstructed terraces, the $D(x,y)$ map without global tilt can be easily utilized for image correction (\emph{e. g.}, by means of affine transformations) based on the well-known $7\times 7$ lattice parameters and experimentally measured values. Afterwards, the corrected images can be used for precise determination of distances between features of interest and periodicity of the step array. This can hardly be done using tilted topography images because only a small fraction of the image can be visualized after contrast level adjustments while some image areas remain invisible. For example, the atomic structure of vicinal Si(5\,5\,7) surface with regular array of triple steps was disputed in literature \cite{Kirakosian-APL-01, Teys-06, Zhachuk-09, Oh-08, Chaika-APL-09} although the high-quality topography images were acquired by different groups. Using the atomically-resolved differential map of the regular triple step array on Si(5\,5\,7) wafers (Fig.~\ref{Fig-02}) we remove uncontrollable STM image distortions using the well-known parameters of the $7\times7$ unit cell. An example of the affine-corrected $D(x,y)$ map obtained for the Si(5\,5\,7) surface with triple step array is presented on Fig.~\ref{Fig-05-Chaika}. The left part of this image shows rather good coincidence between the adatom positions on the $7\times 7$ reconstructed terrace observed experimentally and $7\times 7$ model structure \cite{Oura-03}. At the same time, the right part of the image in Fig.~\ref{Fig-05-Chaika} demonstrates the presence of $5\times 5$--reconstruction on the flat terrace between triple steps \cite{Chaika-APL-09}. The $5\times 5$ model \cite{Oura-03} overlaid onto the right part of the $D(x,y)$ map also shows coincidence of the experimentally measured features with the model atomic structure. Using this map, we define the periodicity of the Si(557) triple step array from either left or right part of the image having different step structures and terrace reconstructions. The distance between neighboring grooves is the same for both structures 'Si(1\,1\,1)$7\times 7$ terrace + triple step' and 'Si(1\,1\,1)$5\times 5$ terrace + triple step' with precision of one interatomic-row distance in the direction perpendicular to the step edges.

\subsection{Removal of a background signal associated with global tilt of the sample}

The procedure of semi-automatic removal of the global tilt described below consists of three stages. The program is written in Matlab programming language, typical execution time is about 10 s.

    \begin{figure*}[t!]
    \centering
    \includegraphics[height=105 mm]{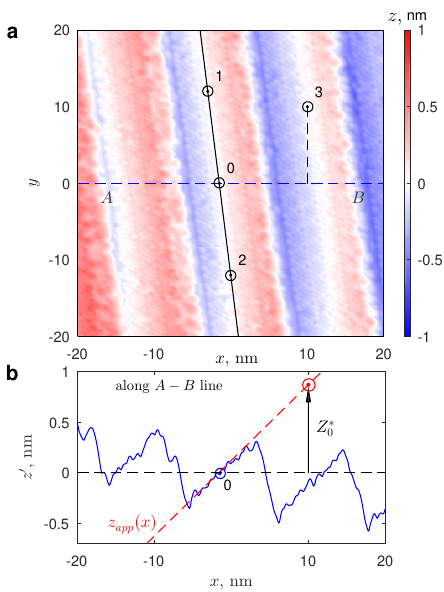}
    \includegraphics[height=105 mm]{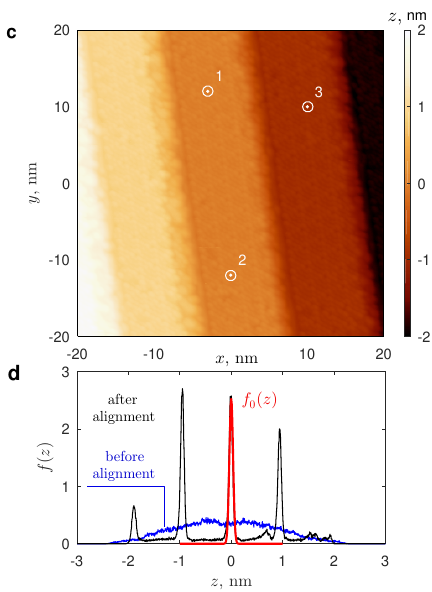}
    \includegraphics[width=155 mm]{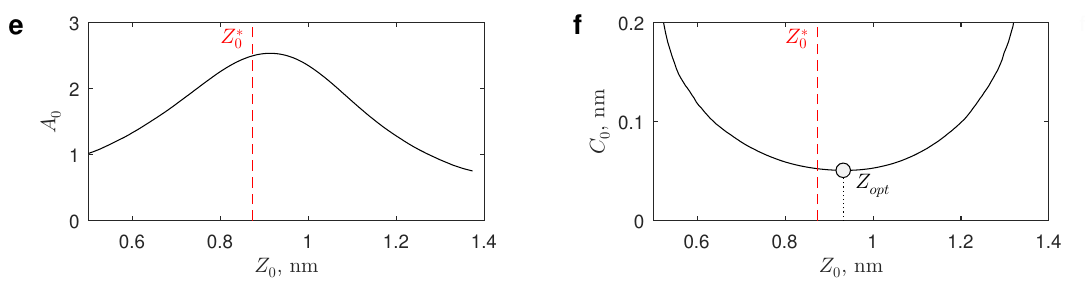}
    \caption{\textbf{a} -- Tilted topography image $z'(x,y)$ of the Si(5\,5\,6) surface (\mbox{$40\times 40\,$nm$^2$}, $U=-0.60\,$V, $I=80\,$pA) after the subtraction of the plane $z^{\,}_p = a^{\,}_0 x+b^{\,}_0 y + c^{\,}_0$, running through the reference points 1, 2 and 3 (the same image is shown in Fig.~\ref{Fig-03}a). The point~0 is the center of the line segment $1-2$. Red-white-blue color map helps us to confirm that the $1-2$ line after the plane subtraction indeed belongs to the plane $z'(x,y)\simeq 0$.  \textbf{b} -- Cross-sectional view of the tilted surface $z'(x,y)$ along the $A-B$ line. Red dashed line is the linear approximation of the dependence $z'(x)$ in the vicinity of the middle point $x^{\,}_0$ [see Eq.~(\ref{Eq-approx-plane-3})]. \textbf{c} -- Aligned topographic image $z^{\prime\prime}(x,y)$ prepared in accordance with Eqs.~(\ref{Eq-approx-plane-0}) and (\ref{Eq-Aligned-Image}). \textbf{d} -- Normalized probability density functions $f(z)$, characterizing the distributions of visible heights for the tilted image $z'(x,y)$ (panel a) and for the aligned image $z^{\prime\prime}(x,y)$ (panel c). Red solid line corresponds to the Gaussian approximation of the zero peak in the probability density function according to the formula (\ref{Eq-gauss-function}). \textbf{e, f} -- Typical dependences of the amplitude $A^{\,}_0$ (panel e) and the width $C^{\,}_0$ (panel f) of the zeroth peak on $Z^{\,}_0$. The optimal value $Z^{\,}_{opt}$ ($\circ$) corresponding to the minimal width of the zeroth peak is close to the estimate of $Z_0^*$ (panel c).
    \label{Fig-06-red-blue}}
    \end{figure*}

At the first stage we determine three reference points (\emph{e. g.}, by clicks of a mouse) in such a way that the points 1 and 2 should be positioned at the same atomically-flat terrace, while the point 3 can be located at any place excluding the line $1-2$ (see Fig.~\ref{Fig-06-red-blue}a). Using the coordinates of the reference points $(x^{\,}_1, y^{\,}_1, z^{\,}_1)$, $(x^{\,}_2, y^{\,}_2, z^{\,}_2)$ and $(x^{\,}_3, y^{\,}_3, z^{\,}_3)$ one can calculate the parameters of the plane
\begin{eqnarray}
\label{Eq-approx-plane-0}
z^{\,}_{p} = a x+b y + c,
\end{eqnarray}
passing through these points, where
\begin{multline}
\label{Eq-approx-plane-01}
a = \frac{(z^{\,}_1-z^{\,}_3)\cdot(y^{\,}_2-y^{\,}_3) - (z^{\,}_2-z^{\,}_3)\cdot(y^{\,}_1-y^{\,}_3)}{(x^{\,}_1-x^{\,}_3)\cdot(y^{\,}_2-y^{\,}_3) - (x^{\,}_2-x^{\,}_3)\cdot(y^{\,}_1-y^{\,}_3)}, \\
b = \frac{(z^{\,}_2-z^{\,}_3)\cdot(x^{\,}_1-x^{\,}_3) - (z^{\,}_1-z^{\,}_3)\cdot(x^{\,}_2-x^{\,}_3)}{(x^{\,}_1-x^{\,}_3)\cdot(y^{\,}_2-y^{\,}_3)-(x^{\,}_2-x^{\,}_3)\cdot(y^{\,}_1-y^{\,}_3)}, \quad \mbox{and} \quad
c = z^{\,}_3 - a^{\,}_0 x^{\,}_3 - b^{\,}_0 y^{\,}_3.
\end{multline}
The result of the subtraction of the plane (\ref{Eq-approx-plane-0}) from raw topography image (Fig.~\ref{Fig-03}a)
\begin{eqnarray}
\label{Eq-approx-plane-02}
z'(x,y) = z(x,y) - \big(a x+b y + c\big),
\end{eqnarray}
is shown in Fig.~\ref{Fig-06-red-blue}a. We will refer to $z'(x,y)$ as tilted topography image. If there are no small-scale corrugation, noise and outliers, the $1-2$ line for the tilted image apparently belongs to the plane $z'(x,y)\simeq 0$. It is important to note that  the procedure (\ref{Eq-approx-plane-01})--(\ref{Eq-approx-plane-02}) minimizes the global slope in the direction along the atomically-flat terraces.

    \begin{figure}[t!]
    \centering
    \includegraphics[height=105 mm]{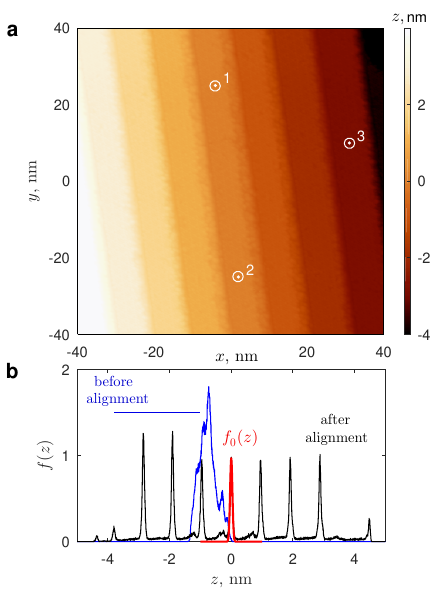}
    \caption{ \textbf{a} -- Aligned topography image for the Si(5\,5\,6) surface  (image size of \mbox{$80\times 80\,$nm$^2$}, $U=-0.60\,$V, $I=80\,$pA) prepared in accordance with Eqs.~(\ref{Eq-approx-plane-0}) and (\ref{Eq-Aligned-Image}). \textbf{b} -- Normalized probability density functions $f(z)$, characterizing the distributions of visible heights for raw image and for aligned image. Red solid line corresponds to the Gaussian approximation of the zero peak in the probability density function according to the formula (\ref{Eq-gauss-function}).
    \label{Fig-07-alignment}}
    \end{figure}

At the second stage we estimate an offset, what is needed to shift the reference point $z^{\,}_3$ in the vertical direction in order to get an aligned topography image with horizontal terraces. We define the point 0 as a middle point of the line segment $1-2$ and then plot the profile of the tilted surface $z'(x, y^{\,}_0)$ along the fast-scanning direction (\mbox{\emph{i. e.}} along the $x-$axis) through this point (Fig.~\ref{Fig-06-red-blue}b). We introduce local linear extrapolation
\begin{eqnarray}
\label{Eq-approx-plane-3}
z^{\,}_{app}(x) = \left(\frac{dz'}{dx}\right)^{\,}_{x^{\,}_0, y^{\,}_0}\cdot (x-x^{\,}_0).
\end{eqnarray}
The dependence of $z^{\,}_{app}(x)$ is shown in Fig.~\ref{Fig-06-red-blue}b by dashed line and it evidences for the pronounced global tilt of the topography image along the $x-$axis at the angle of 4.3$^{\circ}$. We can define the auxiliary parameter $Z^*_0$ as the difference between the extrapolated value for the dependence $z^{\,}_{app}(x)$ at the reference point 3 and zero level (red dot in Fig.~\ref{Fig-06-red-blue}b). It is clear that the repeated subtraction of a plane defined by the points $(x^{\,}_1, y^{\,}_1, 0)$, $(x^{\,}_2, y^{\,}_2, 0)$ and $(x^{\,}_3, y^{\,}_3, -Z^*_0)$ from the titled image $z'(x,y)$, prepared at the first stage, is able to return us a topography image without noticeable slope in the $x-$direction. Unfortunately, this method seems to be ineffective for noisy topography images containing atomically-flat terraces of nanometer-size widths because of substantial uncertainty in estimating the position of the offset $Z^{*}_0$.

In order to overcome this issue, at the third stage we numerically solve single-parameter optimization problem:

(i) for each $Z^{\,}_0$ value, taken in the vicinity of  $Z^*_0$, one can introduce a trial plane defined by the points $(x^{\,}_1, y^{\,}_1, 0)$, $(x^{\,}_2, y^{\,}_2, 0)$, and \mbox{$(x^{\,}_3, y^{\,}_3, -Z^{\,}_0)$}.

(ii) after subtraction this trial plane from the tilted topography image prepared at the first stage, one can compose a histogram $f(z)$, which characterizes the distribution of visible heights $z$ for such partly aligned topographical image. The probability distribution function $f(z)$ for properly aligned image should have a series of narrow peaks (see Fig.~\ref{Fig-06-red-blue}d), and one of these peaks should be definitely positioned at $z=0$ (so called zeroth peak). The zeroth peak can be fitted by a Gaussian function
\begin{eqnarray}
\label{Eq-gauss-function}
f^{\,}_0(z) = A^{\,}_0\,\exp\left(-\frac{z^2}{C^2_0}\right),
\end{eqnarray}
where $A^{\,}_0$ and $C^{\,}_0$ are fitting parameters. Similar histogram-based analysis was already applied for precise determination of the height of monatomic steps on Si(1\,1\,1) surface and subsequent fine calibration of a piezo-scanner \cite{Aladyshkin-23}.

(iii) after sweeping $Z^{\,}_0$, one can get the dependences of the amplitude $A^{\,}_0$ and the width $C^{\,}_0$ of the zeroth peak on the $Z^{\,}_0$ parameter Figs.~\ref{Fig-06-red-blue}e  and \ref{Fig-06-red-blue}f, respectively).

We define the optimum offset value $Z^{\,}_{opt}$ as the value $Z^{\,}_0$ at which the zeroth peak has the minimal width. The results of the additional subtraction of the optimal plane running through the points \mbox{($x^{\,}_1, y^{\,}_1, 0$)}, \mbox{($x^{\,}_2, y^{\,}_2, 0$)} and \mbox{($x^{\,}_3, y^{\,}_3, -Z^{\,}_{opt}$)}, from the tilted image $z'(x,y)$ (Fig.~\ref{Fig-06-red-blue}a)
\begin{eqnarray}
\label{Eq-Aligned-Image}
z''(x,y) = z'(x,y) - (a^{*} x + b^{*} y + c^{*}),
\end{eqnarray}
is presented in Fig.~\ref{Fig-06-red-blue}c;
\begin{multline}
\label{Eq-approx-plane-2}
a^{*} = \frac{ Z^{\,}_{opt}\cdot(y^{\,}_1-y^{\,}_2)}{(x^{\,}_1-x^{\,}_3)\cdot(y^{\,}_2-y^{\,}_3) - (x^{\,}_2-x^{\,}_3)\cdot(y^{\,}_1-y^{\,}_3)}, \\
b^{*} = \frac{Z^{\,}_{opt}\cdot(x^{\,}_2-x^{\,}_1)}{(x^{\,}_1-x^{\,}_3)\cdot(y^{\,}_2-y^{\,}_3)-(x^{\,}_2-x^{\,}_3)\cdot(y^{\,}_1-y^{\,}_3)}, \quad  \mbox{and} \quad
c^{*} = -Z^{\,}_{opt} - a^{\,}_1\cdot x^{\,}_3 - b^{\,}_1\cdot y^{\,}_3.
\end{multline}
Thus, we are able to compose the aligned topography image, which has constant (with experimental accuracy) contrast at each atomically-flat terrace. The aligned topography image for larger scanning area of the Si(5\,5\,6) surface is shown in Fig.~\ref{Fig-07-alignment}a.

\begin{figure}[t!]
    \centering
    \includegraphics[width=80 mm]{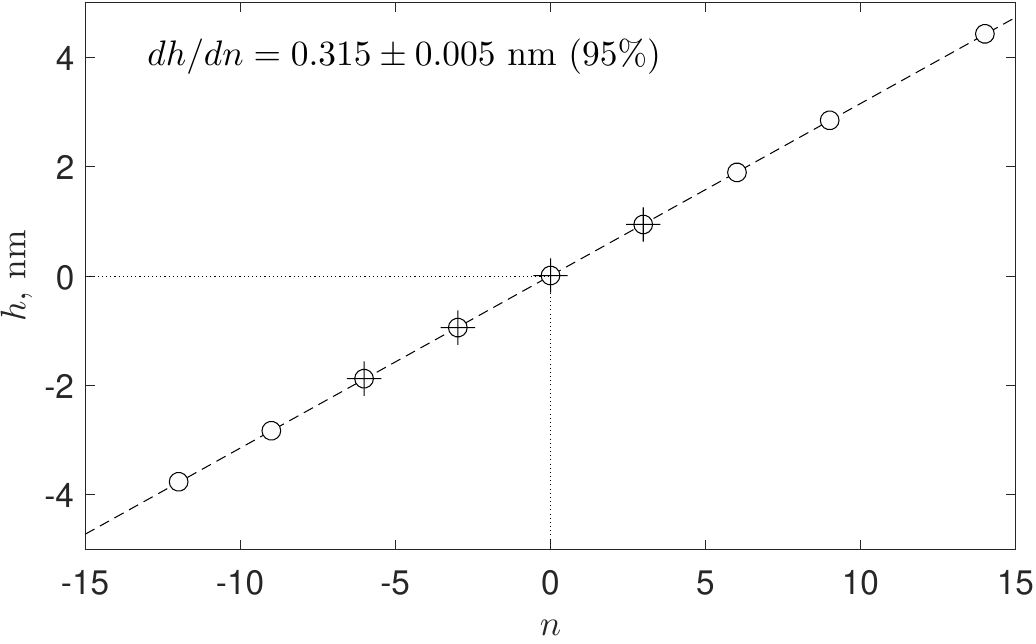}
    \caption{Dependence of the visible height of the atomically-flat terraces $h$ for the optimally aligned Si(5\,5\,6) surface on the number of monolayers $n$ with respect to the terrace defined by the reference points 1 and 2. This plot is composed using the histograms shown in Figs.~\ref{Fig-06-red-blue}d~($+$) and \ref{Fig-07-alignment}b~($\circ$) without any additional calibration. The slope $dh/dn$ is pretty close the to the height of the monolayer for the Si(111) surface ($d_{ML}^{(111)}=0.314$\,nm).
    \label{Fig-08}}
    \end{figure}

As expected, the probability density functions $f(z)$ for these images have series of narrow maxima (Figs.~\ref{Fig-06-red-blue}d and \ref{Fig-07-alignment}b), whose width \mbox{($\sim 0.1\,$nm)} is substantially smaller than the interval between neighbor maxima \mbox{($\sim 0.95\,$nm)}. Statistical analysis of the mutual arrangement of the peaks at these histograms, illustrating relative heights of the multiatomic steps of the aligned Si(5\,5\,6) surfaces, is presented in Fig.~\ref{Fig-08}. We conclude that the most part of the steps at the Si(5\,5\,6) surfaces are triple steps, which have mean height close to the theoretical value $3d^{(111)}_{ML}\simeq 0.941$\,nm, where $d^{(111)}_{ML}=0.3135\,$nm is the thickness of monolayer for the Si(1\,1\,1) surface.

    \begin{figure}[h!]
    \centering
    \includegraphics[width=75 mm]{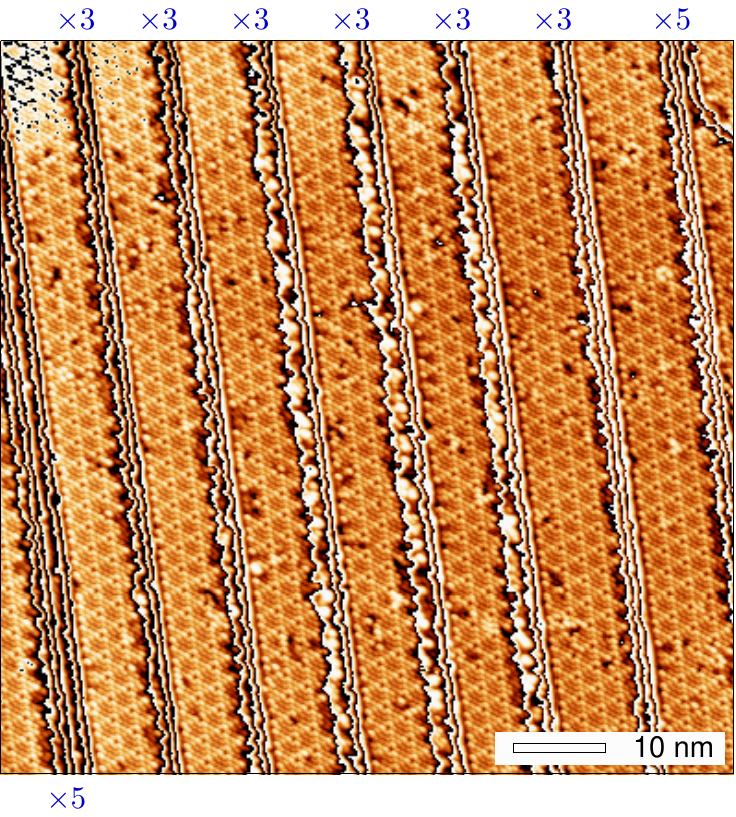}
    \caption{Map of the differential signal $\delta z(x,y)$ after removal of the background signal  [see Eq. (\ref{Eq-no-background})] for the same area of the Si(5\,5\,6) surface as it is shown in Fig.~\ref{Fig-07-alignment}a. Symbols '$\times 3$' and '$\times 5$'  point to the multiplicity factors of corresponding multiatomic steps on the sample surface.
    \label{Fig-09}}
    \end{figure}

The difference of the aligned topographic image $z''(x,y)$ (without global tilt both along $x$- and $y$-directions) and background signal, associated with the series of quantized steps on the sample surface, can be introduced as follows
\begin{equation}
\label{Eq-no-background}
\delta z(x,y) \equiv z''(x,y) - d^{(111)}_{ML}\cdot \left[\frac{z''(x,y)}{d^{(111)}_{ML}}\right],
\end{equation}
where symbol $[\ldots]$ denotes the function which rounds each element in the argument to the nearest integer. Example of applying the procedure (\ref{Eq-no-background}) to the aligned image in Fig.~\ref{Fig-07-alignment}a is shown in  Fig.~\ref{Fig-09}. It is easy to see that this transformation allows to visualize the crystalline lattices at each terrace and determine multiplicity factors for each multiatomic steps at the surface. The latter factor is apparently equal to the number of almost parallel lines running along the steps. Such histogram-based optimization procedure is very useful for experimental studies oriented to the precise determination of the parameters of mono- and multiatomic steps, for example, for the investigation of the bias dependence of the visible height of the monatomic steps in thin Pb(1\,1\,1) films~\cite{Aladyshkin-23}. Similarly to the images obtained using the difference-of-Gaussians approach (\ref{Fig-05-Chaika}), such aligned image can be affine corrected and further utilized for precise determination of the step periodicity and lateral shifts of the surface unit cells in neighboring terraces.

\section{Conclusion}

We report on successful implementation of two simple algorithms allowing us to eliminate global slope of sample and to visualize crystalline structure on vicinal Si(5\,5\,6) and Si(5\,5\,7) surfaces with rather narrow atomically-flat terraces and triple steps. The first method (also known as difference-of-gaussians) is simply remove local slope, associated with large-scale inhomogeneities, from raw topography image. As a result, small-scale corrugations for all tilted terraces become shifted to the plane $z(x,y)\simeq 0$ regardless on actual height and slope of particular terrace, what can be very useful for visualization of lower-lying features in topography images and further detailed analysis by fast Fourier transform. The second method is based on single-parameter optimization problem in seeking the coefficients of the optimal plane, which characterizes the global slope of the sample, by considering the histogram of visible heights and minimizing the width of the peaks on this histogram. After the subtraction of the optimal plane from raw topography image, one can prepare aligned image with horizontal terraces. Further analysis applied to the aligned images makes possible to visualize crystalline structure on vicinal surfaces, precisely define the distances between step edges, and determine multiplicity factor for all multiatomic steps on the sample surface.

\section{Acknowledgements}

The authors are grateful to N.\,V. Abrosimov for providing Si(5\,5\,6) and Si(5\,5\,7) single-crystal wafers and V.\,S.\,Stolyarov for valuable comments. This work was funded by the Russian State Contracts of the Institute for Physics of Microstructures RAS (FFUF-2024-0020, histogram-based analysis) and the Osipyan Institute of Solid State Physics RAS (sample preparation and measurements). A. Yu. Aladyshkin thanks the Ministry of Science and Higher Education of the Russian Federation (project 075-15-2024-632, programming and difference-of-Gaussians analysis).


\begin{references}
\bibitem{Comment}
The angle $\theta$ between two planes, characterized by the Miller indices  \mbox{($h^{\,}_1$, $k^{\,}_1$, $l^{\,}_1$)} and \mbox{($h^{\,}_2$, $k^{\,}_2$, $l^{\,}_2$)}, can be determined from the relationship
\begin{eqnarray}
\label{Eq-angle}
\nonumber
\theta = {\rm arccos}\,\left(\frac{h^{\,}_1h^{\,}_2+k^{\,}_1k^{\,}_2+l^{\,}_1l^{\,}_2}{\sqrt{h^{2}_1+k^{2}_1+l^{2}_1}\,
\sqrt{h^{2}_2+k^{2}_2+l^{2}_2}}\right).
\end{eqnarray}
\end{references}
\end{document}